**Dynamics of Post-disaster Recovery in Behavior-dependent Business Networks**

Chia-Fu Liu[1*], Chia-Wei Hsu[1], and Ali Mostafavi[2]
[1] Ph.D Student, Department of Civil and Environmental Engineering, Texas A&M University, College Station, TX 77843-3136
[2] Professor, Department of Civil and Environmental Engineering, Texas A&M University, College Station, TX 77843-3136

## ABSTRACT

The recovery of businesses in the aftermath of disaster is a critical component of community economic resilience; however, little is known about the underlying dynamics within the network of businesses that shape the speed and spillover effects of recovery. This understanding, based on the underlying processes in the network of businesses, is crucial for characterizing community economic resilience to inform more effective recovery policies and resource allocation. Addressing this gap, this study investigates the extent to which post-disaster business recovery is shaped by the network diffusion process in pre-disaster business dependency networks shaped by visitation behaviors among business points of interest (POI). We developed a network diffusion model to simulate the recovery progression across different businesses in the Louisiana Gulf Coast area and evaluated its performance using empirical data from the aftermath of Hurricane Ida in 2021. The analysis focuses on four key components: (1) the presence of a diffusion process underlying recovery within a network of businesses; (2) the variations in the extent to which the pace of recovery of different business types are dependent on the recovery of other businesses; (3) the specification of recovery multiplier business types whose recovery would accelerate that of a region's overall business network; and (4) the variations in the composition of recovery multipliers in areas with different income levels. The findings reveal that business recovery is governed by diffusion dynamics on these behavior-based dependency networks, with recovery speed closely linked to pre-disaster visitation patterns. We identified retail and service businesses as recovery multipliers whose rapid recovery significantly accelerates overall recovery in the affected region, and thus enhances the economic resilience of the community. Also, the findings reveal that the business types of recovery multipliers vary in high- versus low-income areas. By characterizing the recovery diffusion process, the findings deepen our understanding of the complex network mechanisms shaping economic resilience in communities following disasters and provide important insights for post-disaster business recovery policies and actions.

## 1 Introduction

Community economic resilience is one determinant of the ability of businesses and urban amenities to cope with and recover from disruptions caused by hazardous events [1]. Business recovery is a vital component of the economic resilience of communities [2, 3]; however, the factors that influence the pace of recovery and its ripple effects within networks of businesses in disaster-stricken areas are not well understood. Gaining insight into these dynamics is essential for assessing community economic resilience, as it highlights the underlying processes within business networks that can guide more effective recovery strategies and resource distribution [2].

*Corresponding author: joeyliu0324@tamu.edu

Business recovery refers to resumption of pre-disruption functionality over time [4]. The majority of studies that have examined post-disaster business recovery have focused primarily on factors shaping recovery speed at the individual business level, including the ability to maintain operations, customer retention, pre-disaster financial health, and adaptivity to changing environments [5, 6]. One important, yet under-studied, aspect of business recovery is the role of network effects on the speed and trajectory of recovery across businesses in an impacted region. Recovery of businesses does not occur in isolation [7]; however, little is known about the presence and importance of network ripple effects in post-disaster recovery of businesses. In fact, high connectivity in urban areas enables shocks to ripple across time, space, and various systems. As interconnectivity grows through the underlying socio-spatial networks embedded in communities, the localized recovery of individual businesses becomes connected to the recovery of other businesses. Business closure during hazard events not only directly affects individual amenities but also influences people's daily lives and movement patterns, with the impact spreading to on other businesses and thus delaying overall economic recovery.

Businesses are interconnected through supply chain networks, co-location networks, and behavior-based dependency networks. In particular, behavior-based dependency networks are formed based on human mobility and visitation patterns capturing lifestyle behaviors and shopping patterns that shape business dependencies. Prior studies [8, 9, 10, 11], have shown the importance of behavior-based dependency networks among businesses in delineating urban lifestyle patterns and urban dynamics; however, the role of behavior-based dependency networks in post-disaster business recovery is not fully known. During hazard events, visitation behavior may be temporarily halted. During the post- disaster recovery phase, however, human mobility—an indicator of recovery—tends to follow pre-disaster business dependency networks. In this scenario, a business that recovers late may act as a "blocker," impeding the recovery of the entire business chain. For instance, consider a behavior chain involving a coffee shop, a gasoline station, and a grocery store. If the gasoline station experiences a prolonged recovery, it will hinder mobility from the coffee shop (predecessor) to the grocery store (successor). Thus, the recovery of a point of interest (POI) largely depends on its predecessor in the business dependency network. Conversely, the recovery of a POI within a specific visitation chain facilitates the recovery of the entire chain. Therefore, it is crucial to examine and characterize the underlying diffusion process that governs recovery speed and trajectory in behavior-based dependency networks. This important knowledge gap is mainly due to the limitations of past data and methods to properly capture and represent behavior-based dependency networks to simulate the network processes that shape the spread of recovery in business networks. This limitation can be addressed by using fine-grained anonymized human mobility data to capture, represent, and model behavior-based business dependency networks [12, 11, 13]. For example, a recent study built a behavior-based dependency network using POI-POI mobility data to investigate the spatial cascade of disruptions in business services [11]. The study demonstrated that behavior-based models could better evaluate business resilience during disruption compared to distance-based models. This finding underscores the importance of behavior-based dependency networks, showing the significance of analyzing diffusion processes in behavior-based dependency networks for characterizing post-disaster business recovery and economic resilience.

Accordingly, this study addresses three key questions: (1) To what extent is the recovery of businesses is shaped by the underlying diffusion process in the pre-disaster behavior-based dependency networks? (2) Which business types are most instrumental in contributing to the overall recovery of businesses and economic resilience of communities? (3) In what ways can business recovery efforts be tailored to consider socio-economic status of areas in which businesses operate? To answer these questions and capture the diffusion mechanism of post-disaster recovery via business dependencies, this study implements a social contagion model to evaluate the dynamics of recovery within an urban business system. As shown in Figure 1, this model focuses on the progression of recovery post-disaster, assessing the network diffusion effect. The choice of a network-based dynamic model is rooted in its effectiveness in studying recovery diffusion [14]. In influence-spreading models, certain key nodes (influencers) are activated first, signifying the adoption of a behavior. Inactive nodes then have the potential to become activated based on the influence of active neighbors and other parameters. This process generates influence cascades that sequentially activate subsets of nodes within the network.

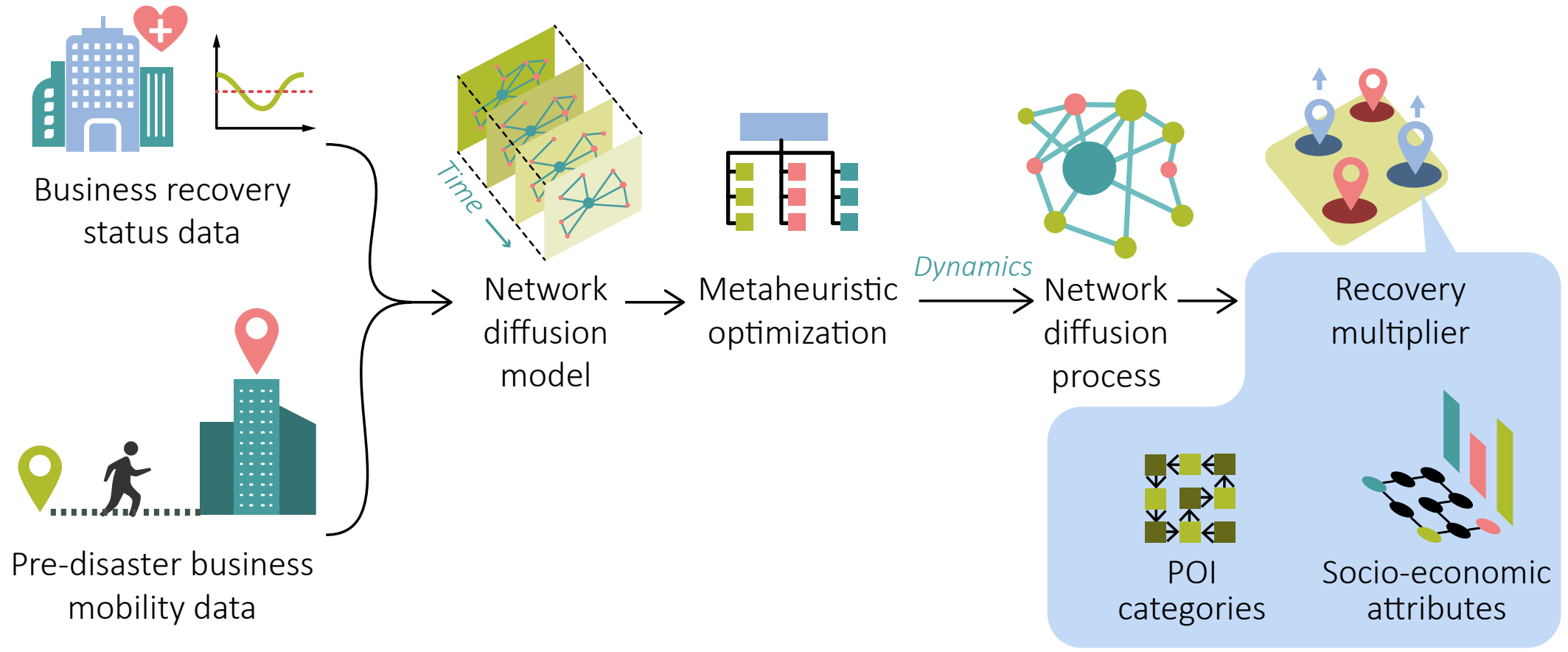

**Figure 1:** Conceptual framework illustrating the two-step optimization process for post-disaster business recovery. The model integrates business recovery status data and pre-disaster mobility data to construct a time-evolving network diffusion model. A metaheuristic optimization is applied to fine-tune the recovery dynamics, transitioning from the network diffusion model to the network diffusion process. This process identifies recovery multipliers, and through geospatial analysis, examines POI categories and socio-economic attributes to guide efficient recovery strategies.

The findings of this study make several significant contributions to our understanding of post-disaster business recovery and economic resilience. First, the findings reveal that post-disaster recovery processes in businesses unfold within behavior-based dependency networks shaped by human mobility patterns. This novel insight highlights the presence of a network diffusion process that influences the speed of business recovery and the broader economic resilience of communities. Also, the findings of this study reveal the importance of human activities and mobility that shape behavior-based dependency networks in post-disaster recovery trajectories of businesses. Second, the network diffusion model and optimization algorithm developed in this study serve as valuable analytical tools for anticipatory analysis, enabling the proactive identification of businesses and areas within the network that are at higher risk of delayed recovery due to their network position. By identifying these isolated businesses and allocating resources to them in advance, the likelihood of permanent business closures, which are often a consequence of insufficiently rapid recovery, can be reduced. Third, the findings highlight the heterogeneity of business types in terms of the dependence on recovery of other businesses. Business sectors such as agriculture, public administration, services, and retail generally demonstrate higher thresholds, indicating a strong interdependence on the recovery of other businesses. In contrast, the wholesale sector, for instance, exhibits lower thresholds, reflecting its capacity to recover more independently. Specifying and prioritizing businesses with high dependency thresholds enables proactive resource allocation to prevent these businesses from becoming isolated and experiencing delayed recovery. Fourth, the study introduces the concept of recovery multiplier businesses—those whose faster recovery accelerates the overall recovery of the regional business network and enhances economic resilience. Prioritizing resource allocation to these key businesses can activate network diffusion effects, thereby speeding up the recovery of the entire network. Finally, the findings uncover subtle differences in the types of recovery multiplier businesses between low- and high-income areas, with retail businesses playing a central role in high-income areas, while service businesses are more crucial in low-income areas. These insights are critical for designing economic recovery plans that promote equitable recovery across all segments of a community. Overall, These outcomes provide interdisciplinary researchers across the fields of disasters, civil engineering, urban science, and geography with a fresh and innovative perspective on the network mechanisms underlying post-disaster business recovery and the economic resilience of affected communities. From a practical perspective, the study outcomes inform policies and practices of emergency management and economic development organizations regarding ways to activate network effects to expedite post-disaster business recovery and the economic resilience of affected communities.

## 2 Datasets

### 2.1 Study context

Hurricane Ida, which made landfall in Louisiana in August 2021, was one of the most devastating hurricanes to impact the United States. Its development from a tropical storm in the Caribbean to a major hurricane was rapid, exhibiting a significant increase in strength as it approached the Gulf Coast. Upon making landfall in Louisiana on August 29, 2021, Hurricane Ida reached Category 4 intensity with sustained winds of 150 miles per hour, making it the second-most intense hurricane to strike the state, following Hurricane Katrina [15, 16].

The hurricane caused catastrophic damage across several Louisiana parishes, with severe winds and heavy rainfall causing widespread flooding and property and infrastructure destruction. The event was particularly noted for its extensive rainfall, which inundated communities, severely damaged infrastructure, and led to prolonged power outages across the region. More than one million residents experienced power losses, and the disaster was responsible for at least 91 fatalities across the affected states [17, 18]. The extent of impacts on people and businesses make this hazard event a unique testbed for this study.

The study area of this study, the Louisiana Gulf Coast area, includes the parishes of St. Bernard, St. James, Plaquemines, Jefferson, St. Tammany, St. John the Baptist, Orleans, and St. Charles in Louisiana. These parishes, equivalent to counties in other U.S. states, represent a diverse cross-section of the region's geographic and socioeconomic landscape. Weekly points-of-interest visits, which serve as a proxy for business activity and community mobility, are collected to form the pre-disaster business dependency network and the business recovery status table to create baseline from which to monitor the recovery progress.

### 2.2 Visitation data and recovery status

The primary aim of this study is to explore the extent to which network diffusion processes within pre-disaster business dependency networks govern business recovery after a disaster. In this context, we examine network models where POIs serve as nodes, encompassing a diverse array of locations such as businesses, public facilities, and infrastructure elements. The interactions or visits between these POIs form the edges of the networks, providing a structural basis for our analysis. In this study, we incorporated 13,017 POIs with 2,414,021 visits among them, resulting in 47,940 visitation paths.

Points of interest represent businesses of different types which are central to our study. These POIs are identified by analyzing the spatial relationships between visitation coordinates and corresponding building polygons which contain detailed place information such as business types and names. This information is derived from a POI table provided by Spectus, our data partner. To ensure the accuracy of POI identification, different buffer zones are tested to optimally assign Standard Industrial Classification (SIC) codes to most coordinates based on proximity to these polygons. The main categories of SIC include retail, finance, services, manufacturing, transport, wholesale, public administration, agriculture, construction, and mining. The geographic distribution of POIs in the Louisiana Gulf Coast area is shown in Figure 2.

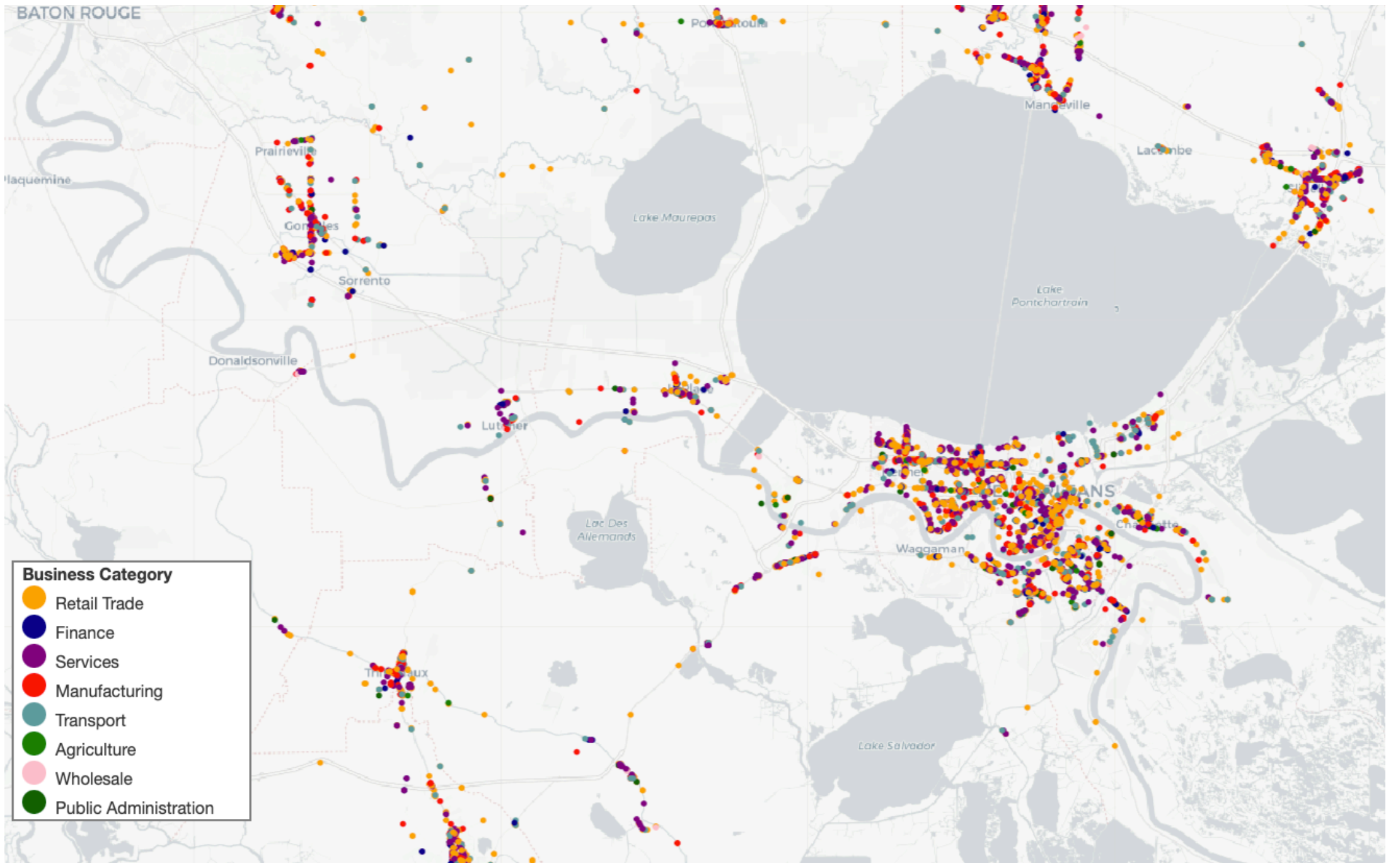

**Figure 2:** Geographic distribution of 13,307 points of interest in the Louisiana Gulf Coast area. The composition of business types is as follows: retail (42.78%), services (16.86%), manufacturing (13.54%), transport (11.29%), finance (10.20%), wholesale (2.89%), public administration (1.85%), and agriculture (0.59%).

Visitation data are sourced from Spectus, a location intelligence data company that provides high spatial accuracy Global Positioning System (GPS) data. This dataset includes an average of 100 data points daily from each anonymized device, encompassing 15 million daily active users in the United States. The collection of this data abides by the General Data Protection Regulation (GDPR) and the California Consumer Privacy Act (CCPA), with all participants providing informed consent for anonymized data collection for research purposes. Previous studies validated the representativeness of the data provided by Spectus [19, 20]. Our data collection comprises daily baseline POI-POI network data for 7 weeks before the disaster (July 1, 2021 to August 18, 2021) to capture normal weekday and weekend visitation patterns, and daily visitation data from POIs for 18 weeks during and after the disaster event (August 26, 2021 to December 29, 2021).

The structure of connections and dependencies between POIs under undisturbed condition form the baseline of behavior- based business networks. Post-disaster daily visitation levels are calculated using a 7-day moving average of visitation to POIs. This data is utilized to determine recovery status by comparing post-disaster activity against pre-disaster norms. A POI is deemed recovered if its post-disaster visitation reaches at least 85% of the baseline level for three consecutive days. We establish a binary recovery status table monitoring recovery on a weekly basis. For example, any POI that achieves recovery within 7 days is designated as week 1 recovery.

## 3 Methodology

To investigate the recovery diffusion process, we developed a contagion model by applying a network diffusion process to the pre-disaster business dependency network. This section outlines the steps for constructing the pre-disaster business dependency network and implementing the post-disaster recovery diffusion simulation.

### 3.1 Pre-disaster business dependency network

To assess the dependencies among businesses in the study region, we first constructed a business network graph, denoted as $G = (V, E)$. Here, each node $v \in V$ represents a POI in the studied area, and each directed edge $e \in E$ corresponds to the mobility flow along an origin-destination (OD) path between these POIs. Essentially, $G$ is a directed graph that encapsulates business dependencies, as indicated by mobility flow during normal, non-disrupted periods. Our study incorporates a total of 13,017 POIs and 47,940 edges in the Louisiana Gulf Coast region. A snapshot of the pre-disaster business dependency network is provided in Figure 3.

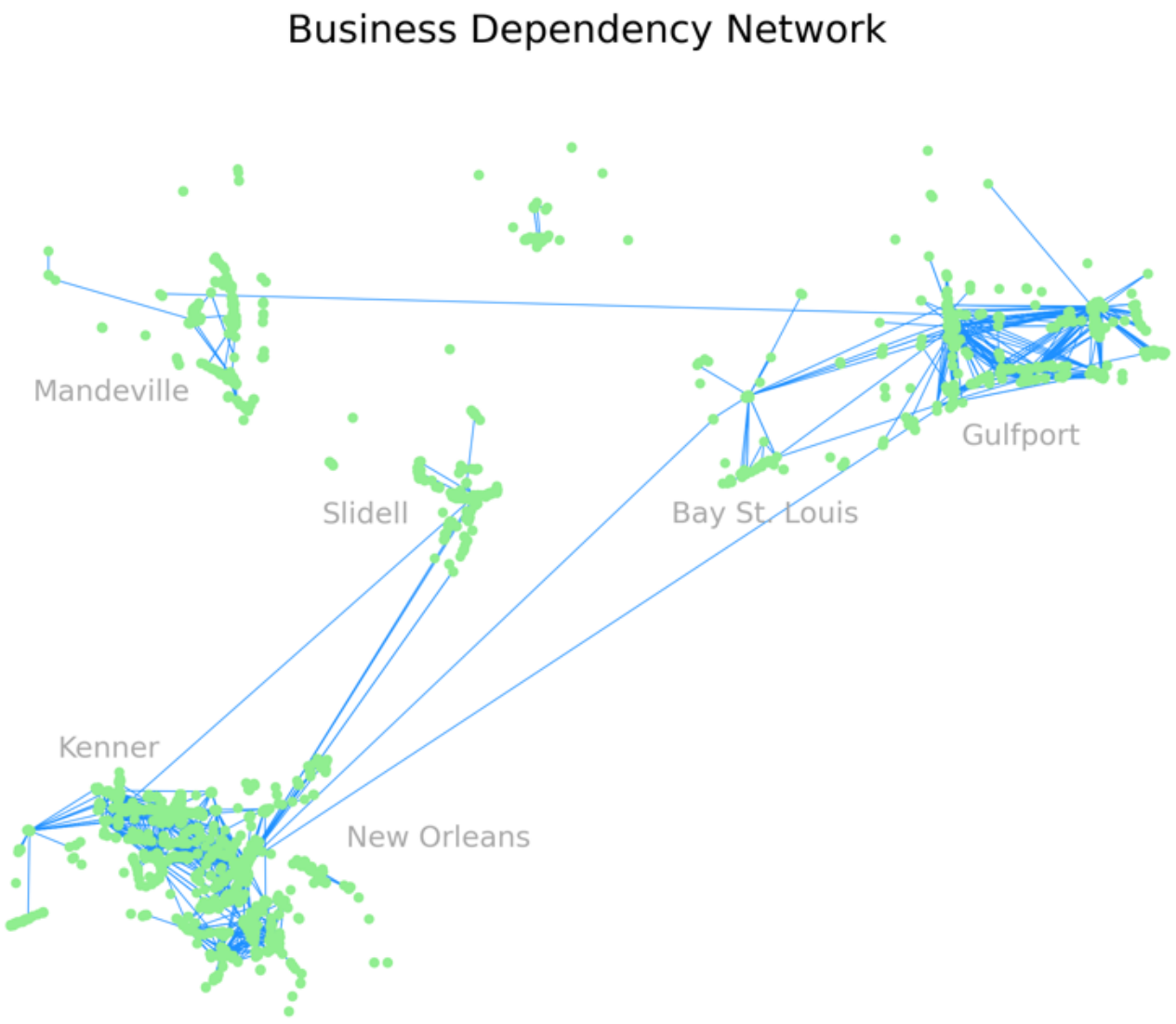

**Figure 3:** Pre-disaster business dependency network in the Louisiana Gulf Coast area, comprising 47,940 visitation paths with an average of 134,112.28 visits per week.

### 3.2 Sub-network construction

To focus on the most significant visitation flow among businesses in constructing the behavior-based dependency network, we filtered sub-networks by applying a lower bound ε for the average count of POI-POI visits. Specifically, only OD paths with an average count exceeding ε were included in the sub-network, based on the rationale that only substantial customer flow is likely to influence a POI's recovery. We set the lower bound ε at the values of (10, 20) to create two sub-networks, I and II, respectively. This approach ensures that only OD mobility flows of significant magnitude are retained. Key characteristics of network graphs are summarized in Table 1. In this context, the order of a graph refers to the number of nodes, while its size refers to the number of edges. Graph density measures the proportion of actual edges relative to possible edges. The vertex degree of a node counts the number of edges connected to it, while vertex strength generalizes degree in weighted networks by summing the weights of edges connected to a particular vertex.

| | Complete netowrk | Sub-network I ($\varepsilon$=10) | Sub-network II ($\varepsilon$=20) |
|---|---|---|---|
| Order | 13,017 | 6,306 | 3,405 |
| Size | 47,940 | 16,738 | 7,091 |
| Density | 5.5728e-4 | 1.123e-3 | 2.095e-3 |
| Transitivity | 0.592 | 0.681 | 0.662 |
| Avg degree | 9.76 | 7.68 | 6.34 |
| Avg strength | 124.33 | 211.48 | 292.62 |

Table 1: Business dependency network graph characteristics

### 3.3 Network diffusion model

To simulate the business recovery progression, we employ the fractional threshold model, a widely recognized social contagion framework [21]. This model is particularly suitable for capturing the dynamics of post-disaster business recovery. Following previous research [14], we model the post-disaster recovery diffusion using the NDlib package [22]. Unlike prior studies where diffusion occurred based on spatial contiguity, our work focuses on the business dependency network, emphasizing how recovery spreads across business networks through behavior-based mobility flows.

The threshold model operates on a straightforward principle: a node becomes active only if the influence exerted by its active neighbors surpasses a certain threshold. In the linear threshold model, a node's influence is calculated as the sum of contributions from its active neighbors, weighted by the strength of their connections. When the cumulative influence exceeds a node-specific threshold, the node becomes active, adopting the recovery behavior. In our study, we implement the fractional threshold model, which considers the fraction of active neighbors rather than their absolute number. The activation condition in this model is

$$\frac{n_i^{act}}{d_i} \geq \theta_i \tag{1}$$

where $n_i^{act}$ is the number of active neighbors of node $i$ and $d_i$ is the degree of node $i$, and $\theta_i$ is the specific threshold of node $i$. A node is activated if the fraction of its active neighbors exceeds its threshold; otherwise, it remains inactive.

The model operates as follows: (1) construct a directed network based on the OD mobility of POIs in the Louisiana Gulf Coast regions. (2) assign a threshold value between 0 and 1 to all nodes. (3) initially activate a set of pre-defined nodes. (4) Perform iterative steps where inactive nodes may become active based on the activation status of their neighbors. During each iteration, all active nodes remain active, while each inactive node is evaluated for activation based on the fraction of its active neighbors meeting or exceeding its threshold. These steps are repeated until the end of the study window or no further nodes can be activated. In our study, an active node represents a POI that has recovered following the hazard event, while an inactive node indicates a POI still struggling to return to pre-disaster normalcy. Each iteration of the model represents a week, with the total study window spanning 18 weeks ($T = 18$).

### 3.4 Recovery diffusion model configuration

This study involves a two-step optimization process. The first step focuses on constructing a post-disaster business recovery diffusion model that accurately reflects the actual recovery process of POIs. The second step involves identifying key recovery multipliers that are critical to the overall recovery process. In the first step, the primary objective of the first step is to determine node-specific thresholds that minimize the discrepancy between the model's predicted recovery status and the observed recovery status. We use the mean absolute error (MAE) as the performance metric to assess the accuracy of the predictions. The objective function is defined as

$$\min_{\theta} MAE = \frac{\sum_{i,t} |S_i(t) - S_i'(t)|}{|V| \times T} \tag{2}$$

where $S_i(t)$ represents the observed state of POI $i$ at week t from the business recovery data, and $S_i'(t)$ is the corresponding state generated by the fractional threshold model. $|V|$ denotes the order of the network graph. Both the observed and model-generated POI statuses are binary, with a value of 0 indicating the POI has not yet recovered and 1 indicating recovery.

We use metaheuristic approached using a genetic algorithm (GA) to solve this non-linear optimization problem. GA is an efficient optimization technique for solving complex problems, known for its ability to avoid local optima [23, 24, 25]. In GA, two key parameters are the population size $M$, which determines the number of solutions considered in each iteration, and the maximum number of iterations $N$ for the search process. While a larger population size increases the likelihood of finding the optimal solution, it also demands more computational resources. In this study, we run the GA optimization with two population sizes $M = (10, 20)$ and set the maximum number of iterations $N = 10,000$.

## 3.5 Business recovery multiplier optimization

This section aims to identify the critical POIs whose expedited recovery would accelerate the overall recovery of the business network. Building on the parameterized fractional threshold model obtained through the GA optimization in the previous section, we simulate the progression of POI recovery and observe how it unfolds within the business dependency network. The strategy is to leverage the structure of the business network to trigger a cascading recovery effect. Specifically, we aim to selectively activate some key POIs that will most effectively propagate recovery through the network.

Our goal is to identify these critical POIs, referred to as recovery multipliers (as described in the previous work [14]), that can unblock and expedite the recovery cascade. The rationale is that by initiating recovery at strategically chosen POIs from the start, the entire business network can achieve full recovery in the shortest possible time. By fortifying an optimal set of POIs within the business dependency network, emergency management and economic development organizations can significantly accelerate the recovery of the entire business ecosystem. To identify the recovery multiplier, we simulate recovery diffusion in both sub-networks I and II using the optimal node-specific thresholds obtained from the model configuration, with a population size of 20. The recovery multipliers are at the beginning of model dynamics. The objective function is formulated as follows

$$\max_{\Omega^\gamma} \sum_i S_i'(T), \gamma \in 0.03, 0.05, 0.1 \tag{3}$$

where the goal is to maximize the number of recovered POIs by the end of the study window ($T = 18$). We consider three scenarios with different sizes of multiplier $\Omega^\gamma$, where $\gamma$ represents 3%, 5%, and 10% of the total studied POIs. We employ the GA to identify the optimal recovery multipliers for these scenarios, using population sizes of 10 and 20, and setting the maximum number of iterations to 5,000.

## 4 Results

Utilizing a unique mobility flow dataset from the Louisiana Gulf Coast area, we construct the business dependency network among the POIs. The analysis of business recovery diffusion during the post-disaster period was conducted through a two-step optimization process. The first step involved configuring the post-disaster business recovery diffusion model; the second step focuses on optimizing the identification of recovery multipliers. The optimization results are presented in this section.

## 4.1 Recovery diffusion model configuration

The goal of the recovery diffusion model configuration is to determine node-specific thresholds that minimize the differ- ence between the model's predicted recovery status and the observed recovery status. The optimization performance of GA in minimizing MAE is shown in Figure 4 below. The Figure 4 (a) and (b) show the GA optimization performance in terms of MAE for sub-networks I and II, respectively. We ran 10,000 iterations to obtain the optimal solution in each setting. The blue dashed line represents the baseline performance, calculated as the mean of 500 randomized solutions. Additionally, we provide the distribution of threshold values in both sub-networks using different optimization algorithm parameters. In both sub-networks I and II, the algorithm achieved smaller MAE values with a larger genome size—20 in our case. Thus, we selected these two sets of threshold values for the next step: recovery multiplier optimization. In Figure 4 (c)-(d), we display the distribution of non-zero threshold values across both sub-networks, as obtained from the GA with two population sizes.

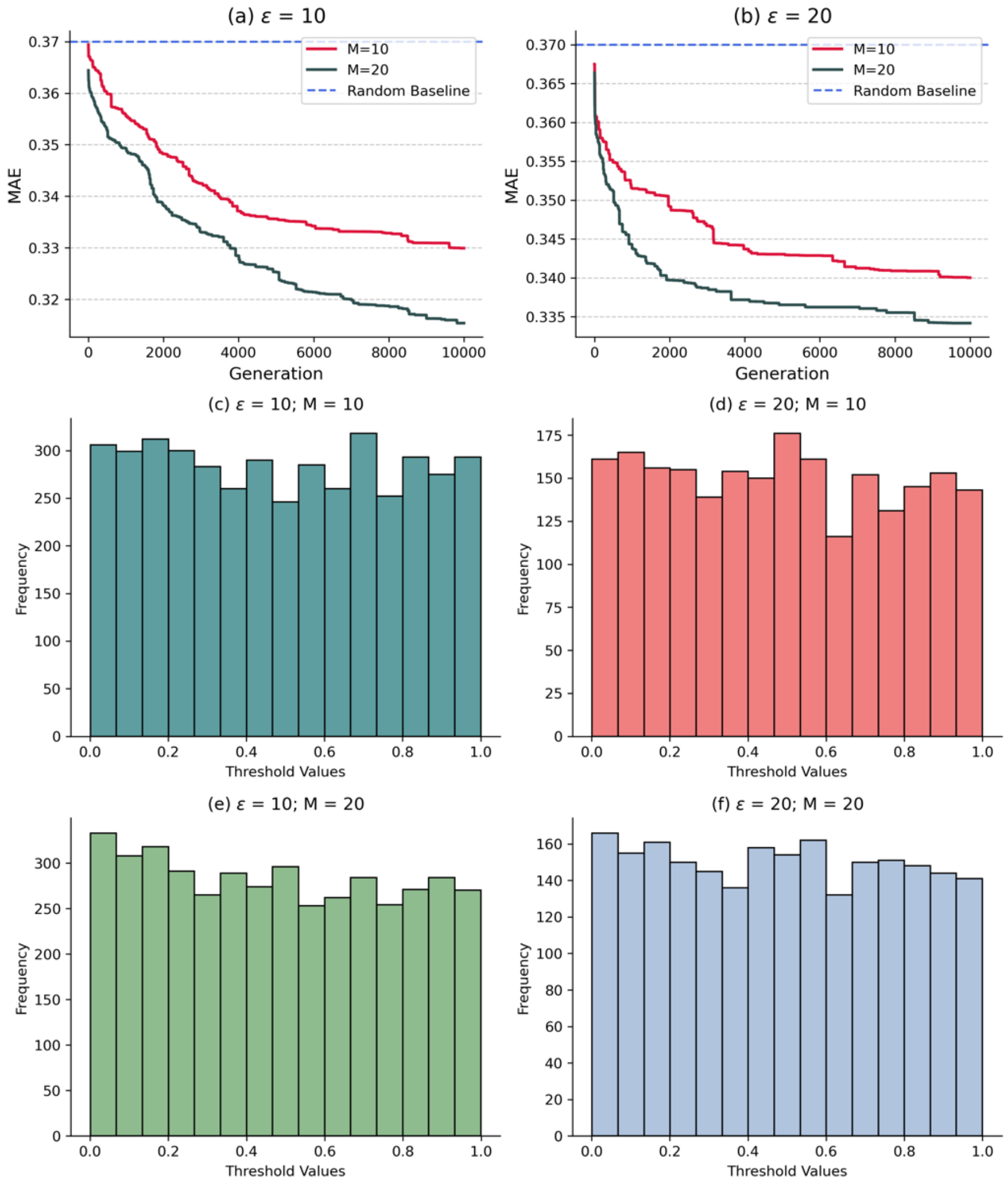

**Figure 4:** (a) GA optimization performance for sub-network I ($\varepsilon = 10$). (b) GA optimization performance for sub-network II ($\varepsilon = 20$). (c) Threshold distribution in sub-network I ($\varepsilon = 10$) using GA algorithm with population size equals 10. (d) Threshold distribution in sub-network II ($\varepsilon = 20$) using GA algorithm with population size equals 10. (e) Threshold distribution in sub-network I ($\varepsilon = 10$) using GA algorithm with population size equals 20. (f) Threshold distribution in sub-network II ($\varepsilon = 20$) using GA algorithm with population size equals 20.

## 4.2 Business recovery multiplier optimization

This section aims to pinpoint the key POIs whose expedited recovery could significantly enhance the overall recovery of the business network. The goal is to maximize the number of recovered POIs by the end of week 18 following Hurricane Ida. As illustrated in Figure 5, we applied the optimal threshold values obtained by setting population size $M$ to 20, as discussed in the previous model configuration section, to determine the recovery multipliers for the two sub-networks. We evaluate three scenarios, each with different sizes of multiplier $\Omega^{\gamma}$, where $\gamma$ corresponds to 3%, 5%, and 10% of the total POIs studied. Specifically, in sub-network I, this equates to 191, 318, and 636 POIs, representing 3%, 5%, and 10% of the graph order. For sub-network II, the corresponding multipliers size are 102, 170, and 340 POIs, representing 3%, 5%, and 10% of the graph order. Figure 6 (a) and (b) show the location of overlapping multipliers among $\Omega^{0.03}$, $\Omega^{0.05}$, and $\Omega^{0.1}$ obtained from the GA optimization with $M = 20$ in sub-network I and II, respectively. There are 37 overlapping multipliers in the sub-network I and 29 in sub-network II. These POIs are of significance as they serve as recovery multipliers, regardless of the $\gamma$ setting.

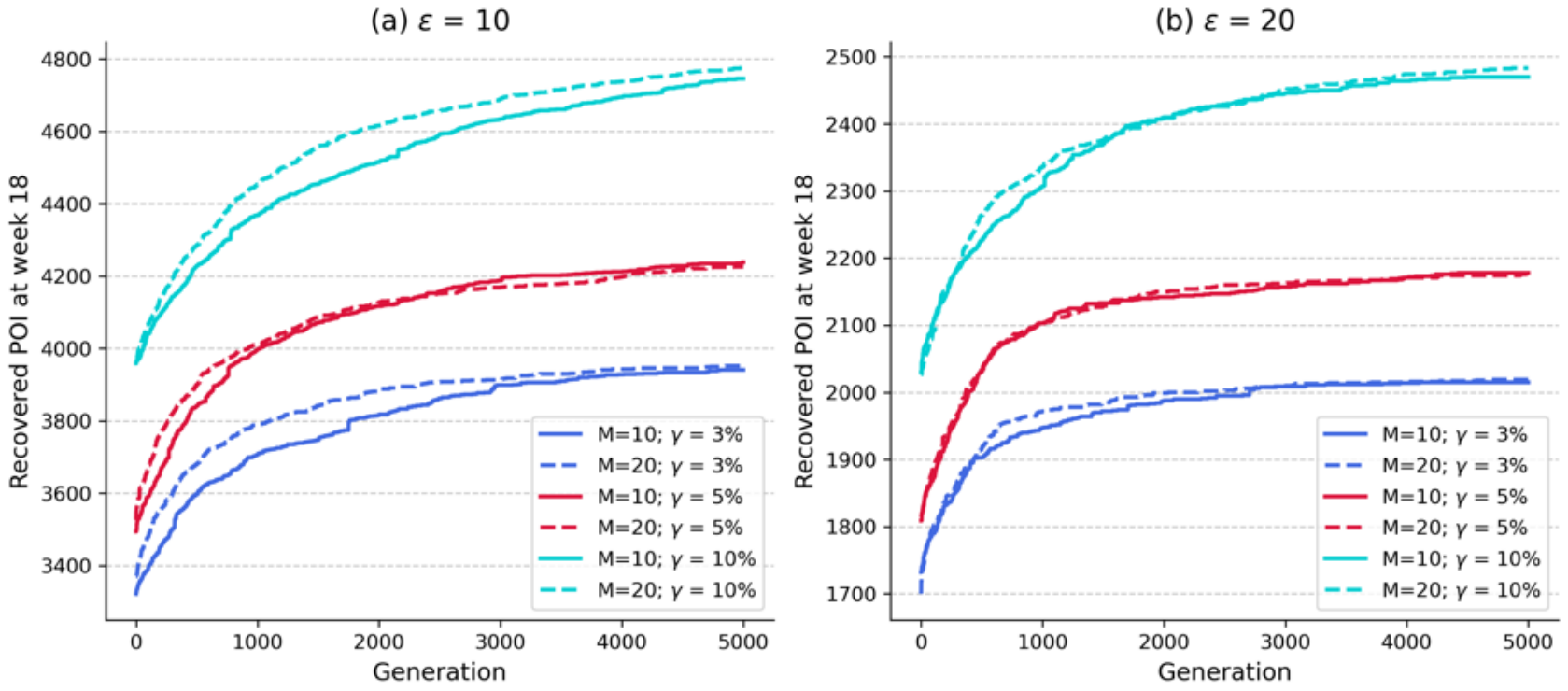

**Figure 5:** (a) GA performance in optimizing recovery multipliers for sub-network I. (b) GA performance in optimizing recovery multipliers for sub-network II.

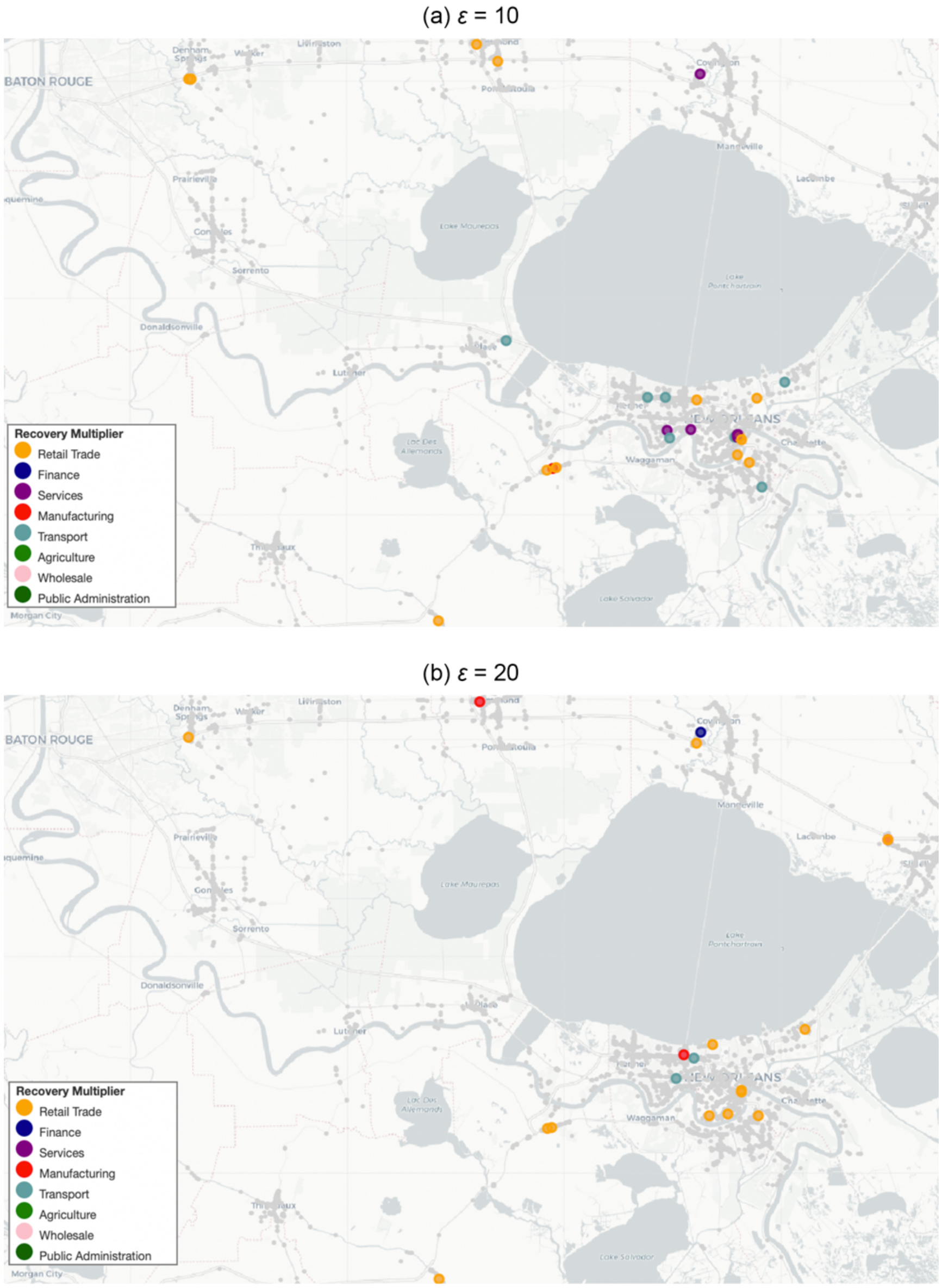

**Figure 6:** (a) Partial view of 37 overlapping recovery multipliers in sub-network I, obtained from GA optimization with $M = 20$. (b) Partial view of 29 overlapping recovery multipliers in sub-network II, obtained from GA optimization with $M = 20$.

## 5 Analysis

### 5.1 Network structure analysis

In this section, we provide an analysis highlighting the structural characteristics of the behavior-based business dependency network. In Figure 7 (a) and (b), the x-axis represents the degree of vertices in the network, plotted on a logarithmic scale. The degree of a vertex, indicating the number of incoming or outgoing edges, reflects the mobility flow pattern within the business dependency network. The steep decline in frequency as the log degree increases suggests that very few nodes possess a high degree, exhibiting a power-law-like distribution typical of scale-free networks. This implies that the majority of businesses are connected to a limited number of other businesses, primarily receiving or sending mobility flow to a small set of POIs. Conversely, a small subset of businesses, which act as major hubs, have a high degree, meaning they are highly connected within the region. These key nodes play a critical role in maintaining the network's structure and connectivity. The network's heavy reliance on these central hubs means that any disruption to them could significantly affect the overall network connectivity and the recovery processes. Therefore, prioritizing these high-degree nodes in disaster recovery efforts could expedite the recovery of the entire network.

Figures 7 (c) and (d) show the vertex strength distribution, which is more normally distributed compared to the degree distribution. Vertex strength measures the total weight of mobility flows that a node sends or receives. The bell-shaped distribution suggests that, despite variations in degree, most POIs have a relatively similar total mobility flow, centered around a specific value (approximately 30). By comparing the degree and strength distributions, it can be inferred that some POIs may lie along limited but strong visitation paths. These nodes have a low degree but higher strength, indicating that they are critical nodes with intense, concentrated mobility flows.

Beyond individual nodes' degree distribution, it is insightful to examine how vertices of different degrees are connected. The scatter plot of average neighbor degree versus vertex degree in Figure 7 (e) and (f) suggests a correlation pattern in which high-degree vertices tend to link with other high-degree vertices, indicating assortative mixing by degree. In contrast, low-degree vertices are connected to both high- and low-degree vertices, showing a more diverse connectivity pattern. This implies that while the network has a core of highly connected nodes, it also exhibits significant interactions between less-connected and more-connected nodes, which could influence recovery dynamics and trajectory.

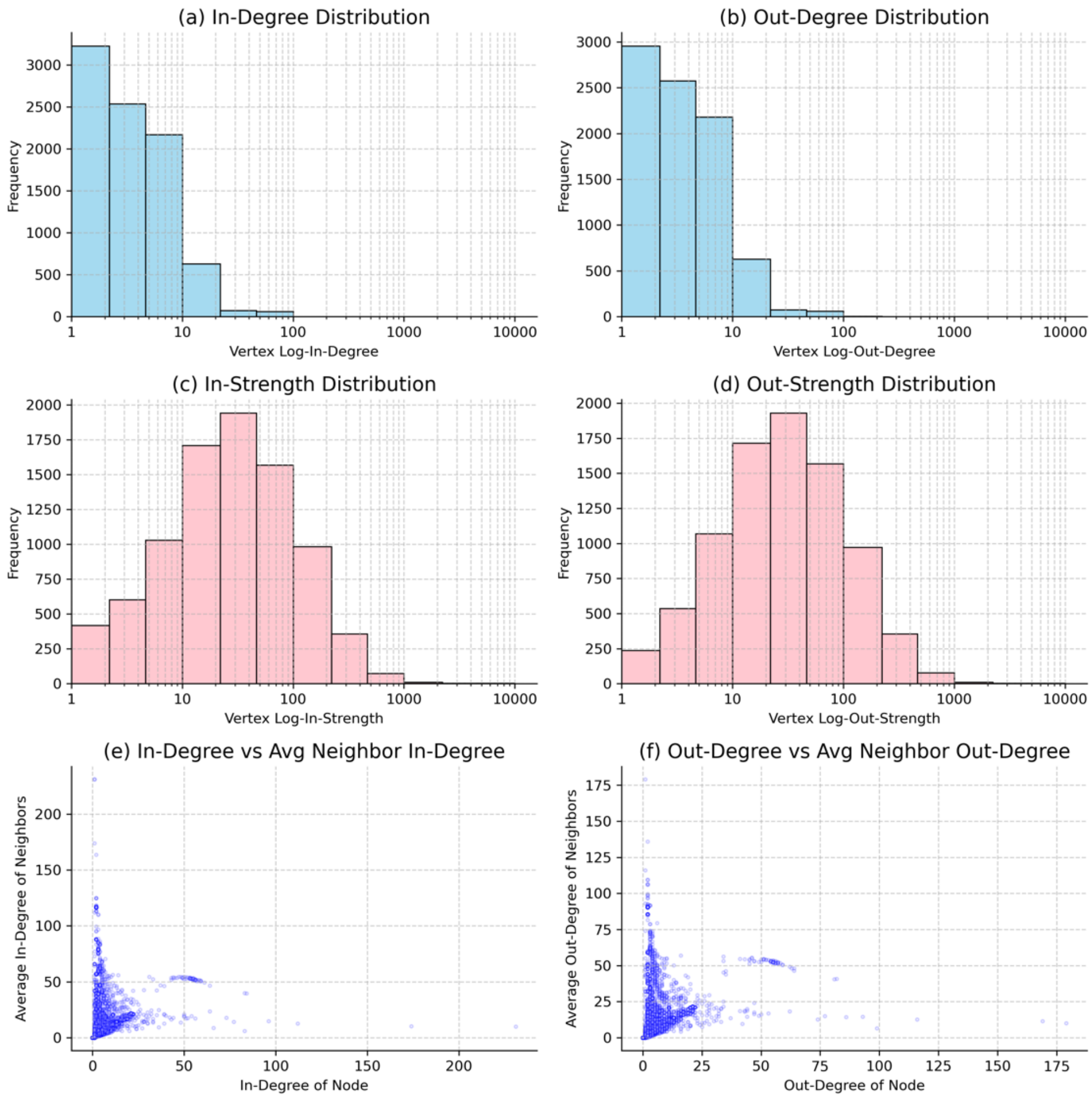

**Figure 7:** (a) In-degree distribution of the business dependency network. (b) Out-degree distribution of the business dependency network. (c) In-strength distribution of the business dependency network. (d) Out-strength distribution of the business dependency network. (e) In-degree vs. average neighbor in-degree in the business dependency network. (f) Out-degree vs. average neighbor out-degree in the business dependency network.

## 5.2 Business recovery threshold analysis

In this section, we delve into different business types' recovery trajectories by examining their recovery thresholds derived from a recovery diffusion model within a POI-POI network. The business types of POIs are classified using the SIC, allowing analysis of the results within and across different sectors.

The focus in this step is the distribution of recovery thresholds among various types of businesses. These thresholds signify the minimal proportion of neighboring businesses that must recover before a specific business can be recovered, an indicator of the extent to which the recovery of a business is dependent on the recovery of other businesses in the network. A higher threshold underscores a business's dependency on other businesses for recovery. Conversely, a lower threshold suggests a business is less dependent on its neighbors.

Figure 8 shows the business-wise and overall threshold distributions and presents central tendencies and threshold variations under different settings (average count lower bound $\varepsilon$ = (10, 20) and genetic algorithm's population size $M$ = (10, 20). Increasing ε sharpens the focus on major businesses by excluding those with lower visitations, highlighting more pronounced differences between business categories. Threshold distributions are consistent across different M, confirming the reliability of our findings. For instance, business sectors such as agriculture, services, and retail generally demonstrate higher thresholds, indicating a strong interdependence on the recovery of other businesses. In contrast, sectors like wholesale and finance exhibit lower thresholds, reflecting their capacity to recover more independently, thus showcasing greater resilience. Furthermore, sectors like services and retail display a higher variability in their recovery thresholds, suggesting a wide range of dependencies within these categories.

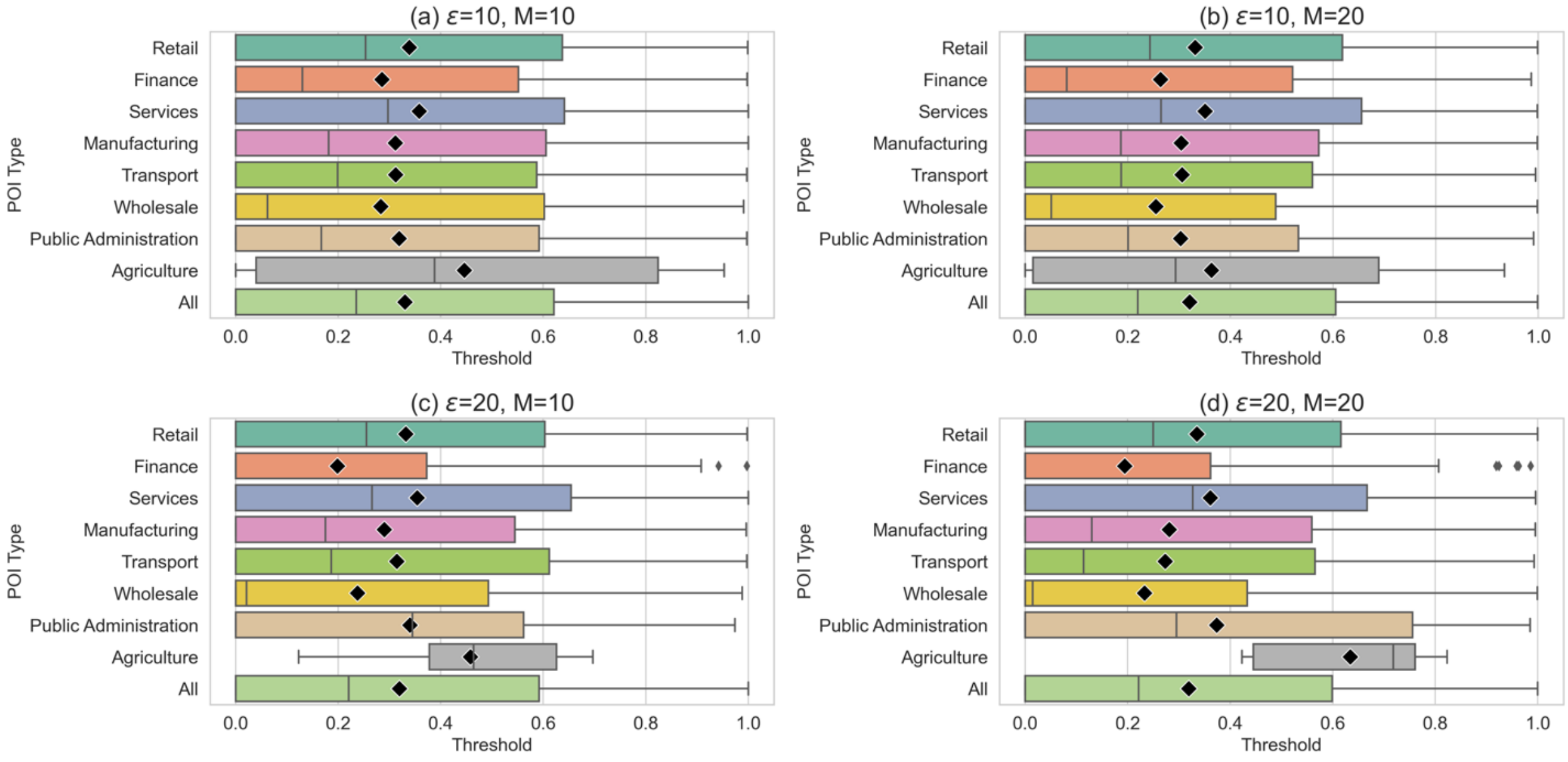

**Figure 8:** Business types threshold distribution across parameter combinations.

We explore the relationship between socio-demographic characteristics, particularly median household income, and the variability of thresholds across different business types. Figures 9 (a), (b), (c), and (d) present the results across parameter settings for each business type. Each figure consists of two subplots: the left subplot displays the overall trend between threshold and income, visualized through a regression line, indicating how threshold values correlate with income levels. The right subplot depicts the distribution of thresholds within high- and low-income areas, determined by the 80th and 20th income percentiles, respectively. Wholesale and public administration businesses situated in lower-income areas tend to exhibit higher threshold values, as indicated by consistently negative slopes across all settings, suggesting a greater dependency on the recovery of neighboring POIs. Conversely, transport and finance in these regions typically demonstrate lower threshold values, with consistently positive slopes, indicating a lesser dependency on neighboring recovery.

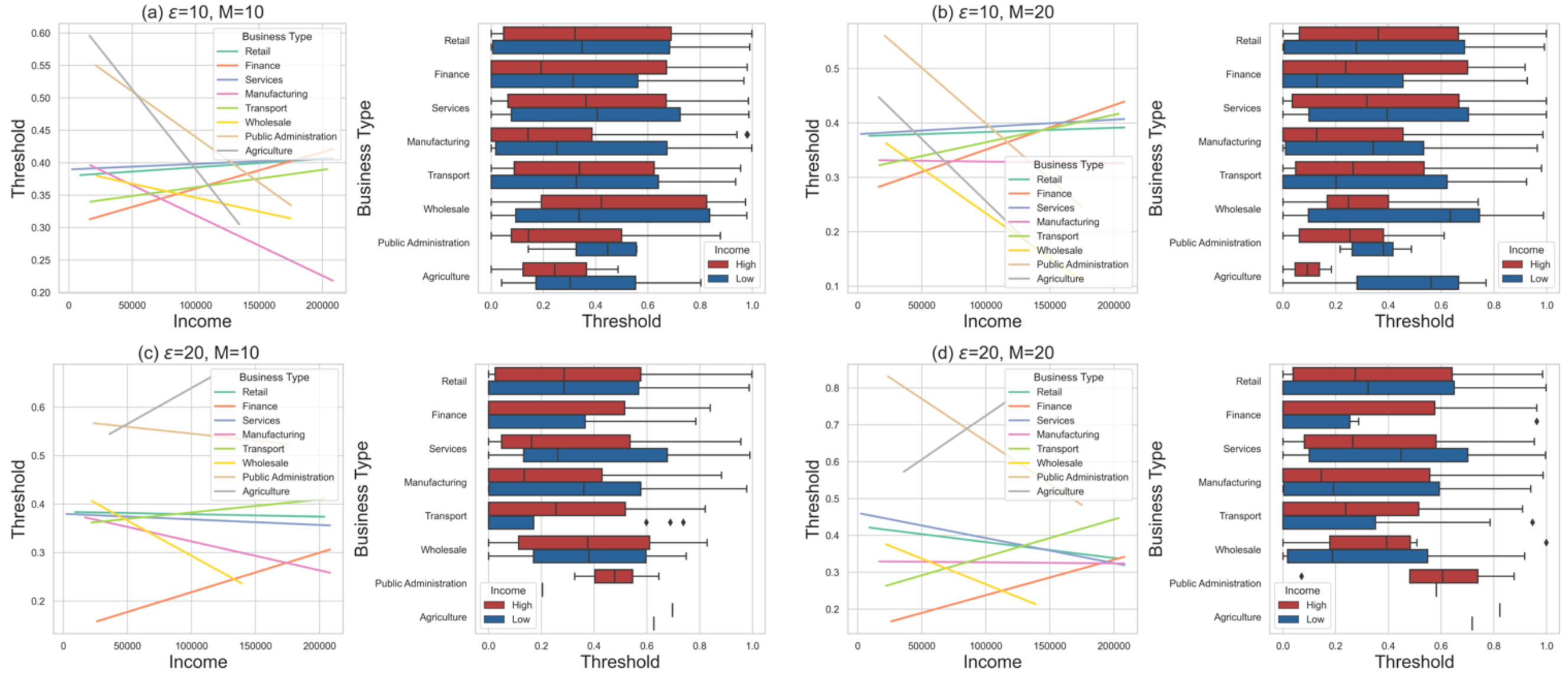

**Figure 9:** Threshold distribution of business types across parameter combinations in high- and low-income areas.

The insights regarding the heterogeneity of different business types in terms of their dependency on other businesses for post-disaster recovery enable identifying businesses with a greater dependency to proactively allocate resources to avoid the creation of recovery isolates (i.e., a cluster of businesses with delayed recovery with a high level of dependency with other businesses).

### 5.3 Business recovery multiplier analysis

In this section, we investigate the role of different business types as potential recovery multipliers within the context of post-disaster recovery scenarios. Our analysis is framed around understanding the business types whose faster (or slower) recovery expedites (or delays) the recovery of the entire business network.

We begin with examining the distribution of business types in our area of interest, highlighting the composition of business types of each category within a functioning POI-POI network. The shares are as follows: retail (42.78%), services (16.86%), manufacturing (13.54%), transport (11.29%), finance (10.20%), wholesale (2.89%), public administration (1.85%), and agriculture (0.59%). This composition provides a baseline for understanding the normal representation of each business type in the landscape of the community.

Next, we analyze the share of each business type identified as recovery multipliers under different scenarios ( $\varepsilon$ = (10, 20), $M$ = (10, 20), $\gamma$ = (3%, 5%, 10%)). The results are shown in Figure 10. This approach allows us to assess whether the contribution of a business type is weighted more or less heavily in facilitating post-disaster business recovery compared to its usual share. When a small number $\gamma$ = 3% of recovery multipliers are selected, retail, services, and wholesale emerge as critical in most scenarios, while manufacturing, transport, and public administration are critical in specific conditions. With a slightly higher threshold γ = 5%, services and wholesale remain critical across most scenarios. Retail, transport, and public administration gain importance in selected scenarios. When a considerable number γ = 10% of recovery multipliers are selected, retail, services, and wholesale consistently stand out as critical. Transport also becomes critical in certain scenarios.

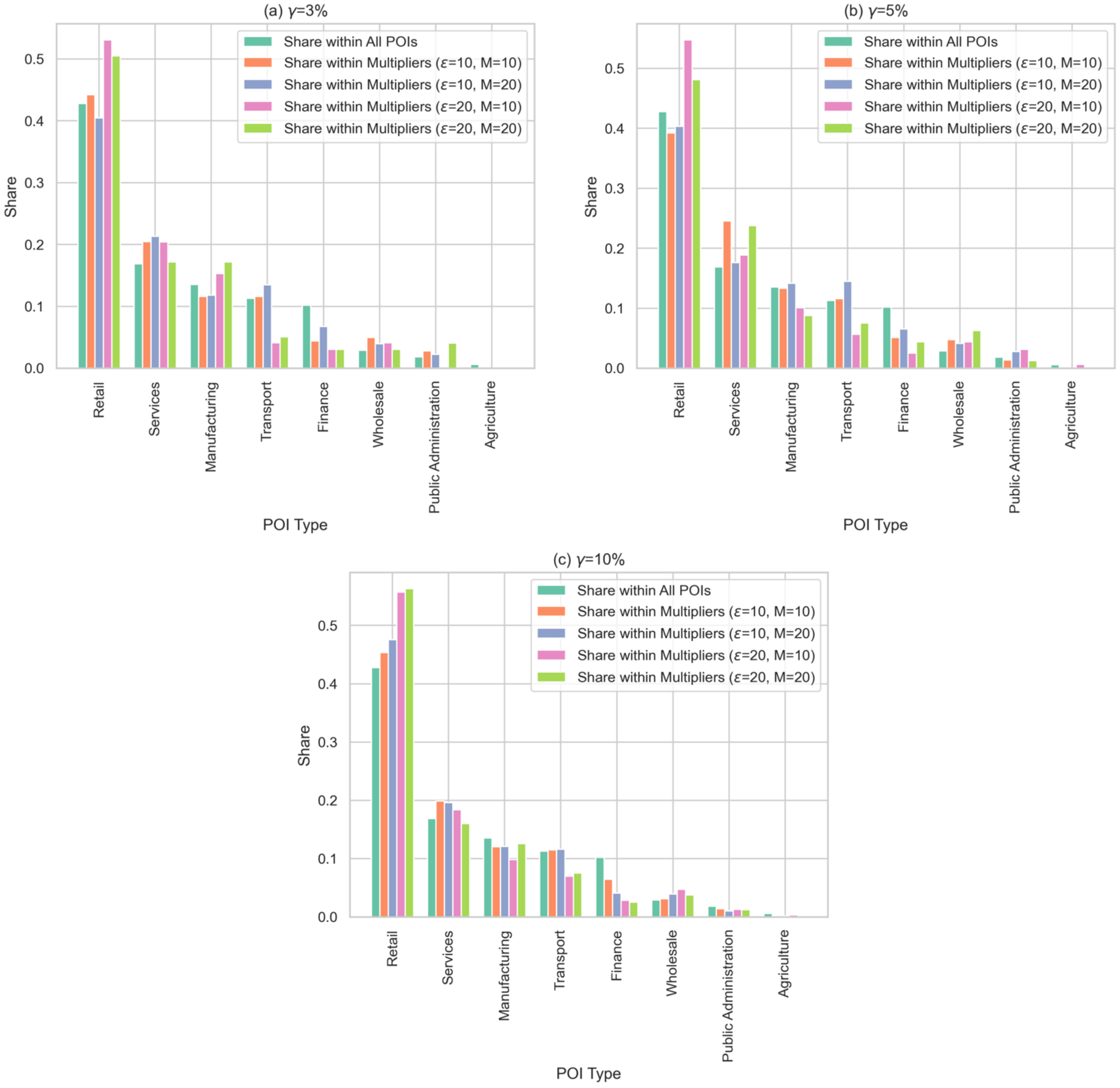

**Figure 10:** Composition of business types identified as recovery multipliers at $\gamma$ = 3%, 5%, and 10%.

Wholesale businesses consistently represent a higher share among recovery multipliers than their representation under normal conditions. Retail and services generally maintain a higher presence among recovery multipliers, underlining their pivotal role in recovery. Conversely, finance and agriculture tend to have lower shares among recovery multipliers, suggesting a lesser significance in contributing the recovery of the business network. Increasing γ values would cause more retail businesses to be identified as multipliers while diminishing the representation of public administration among the multipliers. By comparing the shares of business types under different recovery multiplier number scenarios against their overall representation, the results illuminate the varying degrees to which the recovery of retail and service businesses contributes to the speed of recovery in the overall network of businesses. This understanding can guide strategic planning for more targeted and effective post-disaster recovery resource allocation to the multiplier businesses to leverage network effects to expedite the recovery of all businesses and enhance the economic resilience of the region.

## 5.4 Socio-economic analysis of recovery multipliers

In this part of the analysis, we investigate the consistency of business types identified as recovery multipliers across regions differentiated by socio-economic characteristics, utilizing data from the 2021 American Community Survey (ACS). The ACS provides detailed demographic, social, economic, and housing statistics, serving as a main data source for understanding community conditions and assisting in planning and investment. We categorize census block groups within our area of interest into high-income and low-income regions based on household median income. Specifically, regions falling within the top $20^{th}$ percentile of income are labeled as high-income, whereas those in the bottom $20^{th}$ percentile are classified as low-income.

We record the composition of business types designated as recovery multipliers within these income groups across different selection thresholds ($\gamma = 3\%$, 5%, 10%). As shown in Figure 11, at $\gamma = 3\%$, high-income regions show a clear preference for retail as the primary recovery multiplier, followed by services, manufacturing, finance, transport, wholesale, and public administration. In contrast, low-income regions prioritize services, followed by retail, manufacturing, public administration, and transport. Finance and wholesale do not appear as multipliers at this threshold. When the threshold increases to 5%, the patterns in high-income areas remain generally unchanged, suggesting a consistent recovery strategy that prioritizes similar business types across different settings. However, in low-income areas, while the leading business types—services and retail—remain significant, we observed slight variations in their ranking and the representation of other types. At a threshold of 10%, the differences become more pronounced. In high-income areas, retail and services remain the dominant multipliers, followed by manufacturing and transport, with finance and wholesale also playing roles. Interestingly, in low-income areas, retail surpasses services as the most critical multiplier, and both finance and wholesale are now involved, surpassing public administration. This inclusion reflects a broader and more inclusive recovery dynamic as the threshold widens, suggesting that a wider array of business types could become multipliers under higher $\gamma$ values.

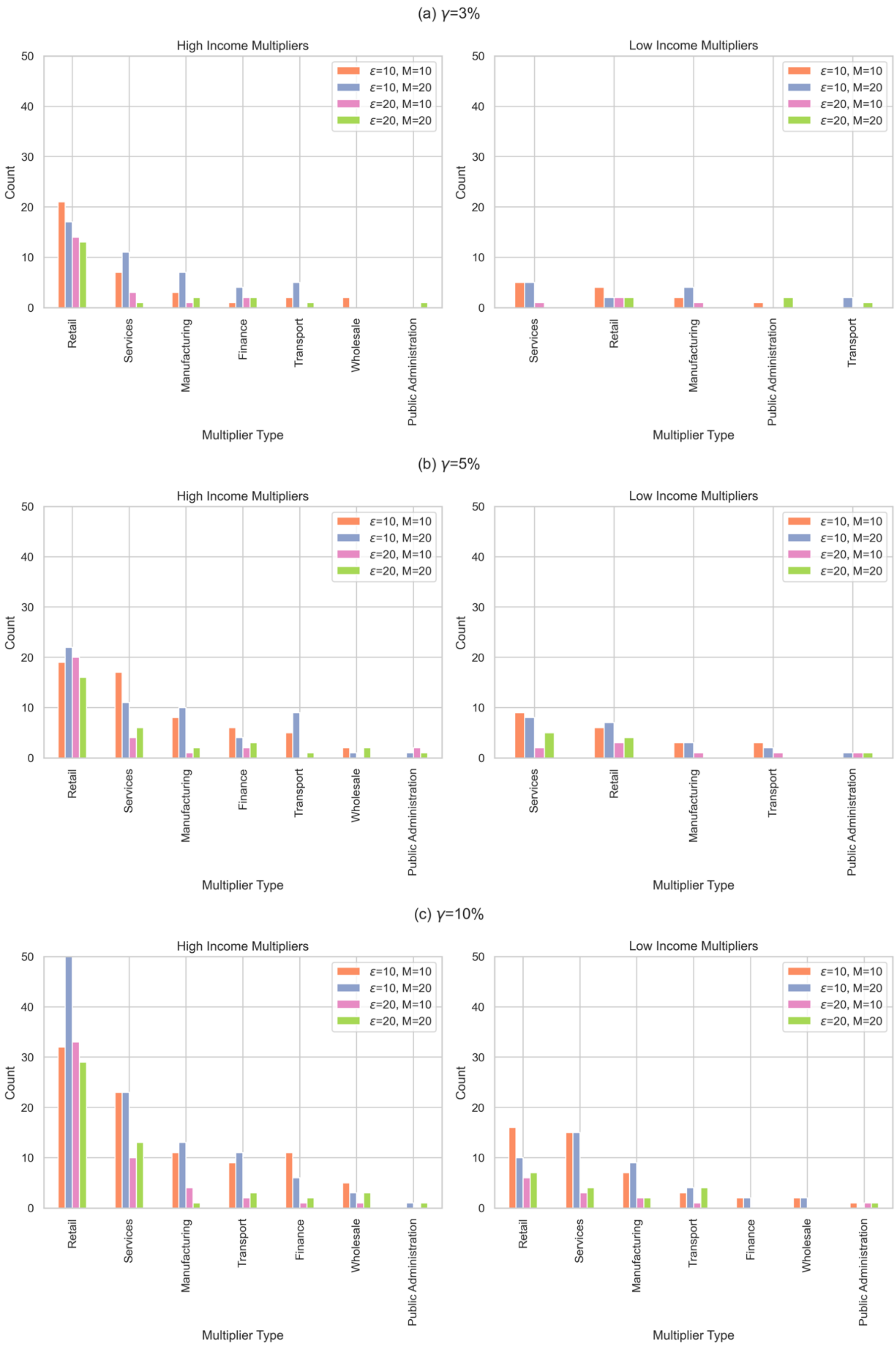

**Figure 11:** Number of business types identified as recovery multipliers in high-income and low-income areas at $\gamma$ = 3%, 5%, and 10%.

The analysis reveals that high-income areas maintain a consistent set of recovery multiplier business types across various thresholds, indicating stable economic structures where certain business types consistently act as key nodes in the recovery process. Conversely, the recovery strategy in low-income areas appears more sensitive to the $\gamma$ values, with a greater variety of business types emerging as critical multipliers as the selection criteria are relaxed. This finding suggests that recovery strategies in economically disadvantaged areas may need to be more flexible and adaptive, considering the changing importance of different business types as recovery progresses.

## 6 Discussion and Concluding Remarks

Post-disaster recovery is a vital component of community resilience. While extensive research has explored factors influencing community recovery [26, 27, 28], relatively less attention has been paid to the recovery of businesses. In particular, the role of business dependency networks in shaping the recovery trajectory of business processes remains rather unexplored. These networks serve as the structural foundation upon which mobility flow and visitation patterns—essential to economic resilience—unfold. In this study, we analyzed business recovery as a network diffusion process, focusing on how behavior-based business dependencies shape recovery across various business types, enabling us to specify the key characteristics of the underlying network diffusion processes to inform strategies for accelerating recovery by leveraging network effects and pinpointing recovery multipliers.

The findings offer significant scientific and practical contributions. Through network structure analysis, we discovered that business dependency networks exhibit a scale-free characteristic. This means that major POI hubs, regardless of their node degree, are densely connected by mobility flow. Protecting and prioritizing these critical business POIs can enhance post-disaster recovery by "unblocking" mobility flow in behavior-based dependent networks, thereby facilitating faster recovery across the entire business network.

The analysis of recovery thresholds across different business types reveals key insights regarding varying levels of dependency. Agriculture, services, and retail sectors show higher recovery thresholds, indicating strong reliance on the recovery of other businesses. Conversely, sectors such as wholesale and finance exhibit lower thresholds, reflecting greater resilience and the ability to recover independent of the recovery status of the connected businesses. The observed variability within sectors, particularly services and retail, suggests heterogeneous dependency levels for these business types. When socio-demographic factors are considered, wholesale and public administration businesses located in lower-income areas tend to have higher threshold values, reflecting a greater dependence on the recovery of neighboring POIs. Conversely, transport and finance in these regions typically demonstrate lower threshold values, indicating less reliance on neighboring recovery. Identifying sectors with higher network dependencies can guide more efficient disaster recovery planning and implementation to leverage network effects in accelerating the recovery of businesses with a greater dependency on the recovery of other businesses.

Our study also highlights the role of different business types as recovery multipliers in post-disaster scenarios, with variation depending on business sector and socio-economic context. Retail, services, and wholesale types consistently emerge as key recovery multipliers, emphasizing their importance in facilitating broader recovery efforts. Interestingly, wholesale businesses, despite their lower representation in business network (in terms of count of POIs), play a disproportionately larger role as recovery multipliers. In contrast, the finance and agriculture sectors appear less critical as recovery multipliers, suggesting that immediate effort may prioritize other sectors to accelerate the recovery of the network of businesses.

Socio-economic factors of locations in which businesses are located further influence recovery dynamics. High-income regions tend to have stable recovery multipliers consisting of retail and services across various multiplier thresholds. In contrast, multiplier business types vary in low-income regions depending on the multiplier number threshold with, a broader array of business types becoming critical as selection criteria expand. This finding underscores the need for adaptable recovery strategies that consider different economic structures of areas in which businesses reside, ensuring efforts are both targeted and equitable across socio-economic landscapes.

In sum, this study contributes to our understanding of post-disaster business recovery and economic resilience in several ways: first, it identifies behavior-based dependency networks, influenced by human mobility patterns, as a spatial structure on whichthe business recovery processes unfold. This suggests that recovery may involve a spatial network diffusion process, highlighting the role of human activities in shaping business recovery trajectories. Second, the study presents a network diffusion model and optimization algorithm that could be useful for anticipatory analysis for business recovery in future disasters. These tools may help identify businesses at higher risk of delayed recovery due to their network position, potentially allowing for more targeted resource allocation. Third, the findings indicate varying levels of interdependence among business types. Agriculture, public administration, services, and retail sectors appear to have higher thresholds, suggesting stronger dependence on other businesses' recovery. Wholesale businesses, in contrast, show lower thresholds, possibly indicating more independent recovery potential. This information could be valuable for prioritizing support to highly dependent businesses. Fourth, the study introduces the concept of "recovery multiplier businesses"—those whose recovery may accelerate the revival of the broader business network. Focusing resources on these businesses could potentially speed up overall recovery. Lastly, the research notes some differences between low- and high-income areas in terms of key recovery businesses. Retail businesses seem to play a central role in high-income areas, while service businesses appear more crucial in low-income areas. This insight could be relevant for developing more tailored economic recovery plans. These findings offer a new perspective on the underlying socio-spatial network mechanisms involved in post-disaster business recovery and community economic resilience. They may be of interest to researchers in fields including disaster studies, civil engineering, urban science, and geography.

The practical implications of this research study could be profound for emergency management and economic development organizations. By uncovering the behavior-based dependency networks that drive post-disaster business recovery, the study provides actionable insights into how human visitation patterns shape behavioral dependency among businesses and influence the speed and trajectory of recovery efforts. The network diffusion model and optimization algorithm developed in the study offer powerful tools for anticipatory analysis, allowing policymakers and public officials to identify and support businesses at higher risk of delayed recovery, thereby reducing the likelihood of permanent closures. The identification of recovery multiplier businesses—those whose quick recovery can accelerate the overall recovery of the regional business network—enables targeted resource allocation, ensuring a faster and more equitable recovery across economic sectors and income levels. These findings can guide the design of more effective and equitable disaster recovery plans, ultimately enhancing the economic resilience of communities.

## Acknowledgments

This material is based in part upon work supported by the National Science Foundation under Grant CMMI-1846069 (CAREER). Any opinions, findings, conclusions, or recommendations expressed in this material are those of the authors and do not necessarily reflect the views of the National Science Foundation. The authors also would like to acknowledge data support from Spectus. Any opinions, findings, conclusions, or recommendations expressed in this material are those of the authors and do not necessarily reflect the views of the National Science Foundation, Texas A&M University, or Spectus.

## Author Contributions

All authors critically revised the manuscript, provided final approval for publication, and agree to be held accountable for the work performed therein. C.F. and C.H., Ph.D. student researchers, conceived the idea, collected the data, performed the analysis, and drafted the manuscript. A.M., the faculty advisor, offered critical feedback on the project's development and manuscript.

## Data availability

All data were collected through a CCPA- and GDPR-compliant framework and utilized for research purposes. The data that support the findings of this study are available from Spectus, but restrictions apply to the availability of these data, which were used under license for the current study. The data can be accessed upon request submitted to the providers. The data was shared under a strict contract through Spectus' academic collaborative program, in which they provide access to de-identified and privacy-enhanced mobility data for academic research. All researchers processed and analyzed the data under a non-disclosure agreement and were obligated not to share data further or to attempt to re-identify data.

## Code availability

The code that supports the findings of this study is available from the corresponding author upon request.

## References

[1] Cristian Podesta, Natalie Coleman, Amir Esmalian, Faxi Yuan, and Ali Mostafavi. Quantifying community resilience based on fluctuations in visits to points-of-interest derived from digital trace data. *Journal of the Royal Society Interface*, 18(177):20210158, 2021.

[2] Stephanie E Chang and Adam Z Rose. Towards a theory of economic recovery from disasters. *International Journal of Mass Emergencies & Disasters*, 30(2):171–181, 2012.

[3] Josephine Adekola and David Clelland. Two sides of the same coin: Business resilience and community resilience. *Journal of Contingencies and Crisis Management*, 28(1):50–60, 2020.

[4] Maria I Marshall and Holly L Schrank. Small business disaster recovery: a research framework. *Natural Hazards*, 72(2):597–616, 2014.

[5] Stephanie E Chang, Charlotte Brown, John Handmer, Jennifer Helgeson, Yoshio Kajitani, Adriana Keating, Ilan Noy, Maria Watson, Sahar Derakhshan, Juri Kim, et al. Business recovery from disasters: Lessons from natural hazards and the covid-19 pandemic. *International Journal of Disaster Risk Reduction*, 80:103191, 2022.

[6] Gary R Webb, Kathleen J Tierney, and James M Dahlhamer. Businesses and disasters: Empirical patterns and unanswered questions. *Natural Hazards Review*, 1(2):83–90, 2000.

[7] Fan Li and Jingke Hong. A spatial correlation analysis of business operating status after an earthquake: a case study from lushan, china. *International Journal of Disaster Risk Reduction*, 36:101108, 2019.

[8] Xue Zhang, Yifan Tang, and Yanwei Chai. Spatiotemporal-behavior-based microsegregation and differentiated community ties of residents with different types of housing in mixed-housing neighborhoods: A case study of fuzhou, china. *Land*, 12(9):1654, 2023.

[9] Junwei Ma and Ali Mostafavi. Decoding the pulse of community during disasters: Resilience analysis based on fluctuations in latent lifestyle signatures within human visitation networks. *arXiv preprint arXiv:2402.15434*, 2024.

[10] Chen Zhong, Stefan Müller Arisona, Xianfeng Huang, Michael Batty, and Gerhard Schmitt. Detecting the dynamics of urban structure through spatial network analysis. *International Journal of Geographical Information Science*, 28(11):2178–2199, 2014.

[11] Takahiro Yabe, Bernardo Garcia-Bulle, Morgan Frank, Alex Pentland, and Esteban Moro. Behavior-based dependency networks between places shape urban economic resilience. 2024.

[12] NatalieColeman,ChenyueLiu,andAliMostafavi.Anatomizingsocietalrecoveryatthemicroscale:Heterogeneity in household lifestyle activities rebounding after disasters. *arXiv preprint arXiv:2406.17993*, 2024.

[13] Junwei Ma, Bo Li, and Ali Mostafavi. Characterizing urban lifestyle signatures using motif properties in network of places. *Environment and Planning B: Urban Analytics and City Science*, 51(4):889–903, 2024.

[14] Chia-Fu Liu and Ali Mostafavi. Network diffusion model reveals recovery multipliers and heterogeneous spatial effects in post-disaster community recovery. *Scientific Reports*, 13(1):19032, 2023.

[15] John L II Beven, Andrew Hagen, and Robbie Berg. National hurricane center tropical cyclone report hurricane ida. *National Hurricane Center*, 2022.

[16] Michael J Hudson and Tony Coventry. Service assessment 2021 hurricane ida. *National Weather Service*, 2023.

[17] Yi-Jie Zhu, Jennifer M Collins, Philip J Klotzbach, and Carl J Schreck III. Hurricane ida (2021): rapid intensifica- tion followed by slow inland decay. *Bulletin of the American Meteorological Society*, 103(10):E2354–E2369, 2022.

[18] Oscar Higuera Roa and Jack O'Connor. Hurricane ida. *United Nations University Institute for Environment and Human Security*, 2022.

[19] Chia-Wei Hsu, Chenyue Liu, Kiet Minh Nguyen, Yu-Heng Chien, and Ali Mostafavi. Do human mobility network analyses produced from different location-based data sources yield similar results across scales? *Computers, Environment and Urban Systems*, 107:102052, 2024.

[20] Chia-Wei Hsu, Chia-Fu Liu, Laura Stearns, Sam Brody, and Ali Mostafavi. Spillover effects of built-environment vulnerability on resilience of businesses in urban crises. *Preprint*, 2024.

[21] Mark Granovetter. Threshold models of collective behavior. *American journal of sociology*, 83(6):1420–1443, 1978.

[22] Giulio Rossetti, Letizia Milli, and Salvatore Rinzivillo. Ndlib: a python library to model and analyze diffusion processes over complex networks. In *Companion Proceedings of the The Web Conference 2018*, pages 183–186, 2018.

[23] Annu Lambora, Kunal Gupta, and Kriti Chopra. Genetic algorithm-a literature review. In *2019 international conference on machine learning, big data, cloud and parallel computing (COMITCon)*, pages 380–384. IEEE, 2019.

[24] Rachid Chelouah and Patrick Siarry. A continuous genetic algorithm designed for the global optimization of multimodal functions. *Journal of Heuristics*, 6:191–213, 2000.

[25] Stefan Etschberger and Andreas Hilbert. Multidimensional scaling and genetic algorithms: A solution approach to avoid local minima. Technical report, Arbeitspapiere zur mathematischen Wirtschaftsforschung, 2002.

[26] Dantje Sina, Alice Yan Chang-Richards, Suzanne Wilkinson, and Regan Potangaroa. What does the future hold for relocated communities post-disaster? factors affecting livelihood resilience. *International journal of disaster risk reduction*, 34:173–183, 2019.

[27] Kathleen Tierney and Anthony Oliver-Smith. Social dimensions of disaster recovery. *International Journal of Mass Emergencies & Disasters*, 30(2):123–146, 2012.

[28] Amber Himes-Cornell, Carlos Ormond, Kristin Hoelting, Natalie C Ban, J Zachary Koehn, Edward H Allison, Eric C Larson, Daniel H Monson, Henry P Huntington, and Thomas A Okey. Factors affecting disaster prepared- ness, response, and recovery using the community capitals framework. *Coastal Management*, 46(5):335–358, 2018.